\documentclass[aps,prd,twocolumn,superscriptaddress,showpacs,amsmath,amssymb]{revtex4}
\usepackage{epsf}
\usepackage{latexsym}
\usepackage{amsmath}
\usepackage{amsfonts}
\usepackage{amssymb}
\usepackage{graphicx}
\newcommand{\be}{\begin{eqnarray}}
\newcommand{\ee}{\end{eqnarray}}
\newcommand{\ud}{\mathrm{d}}

\newcommand{\lp}{\ell_{\rm p}}
\newcommand{\mpl}{m_{\rm p}}

\begin{document}
\title{Point-like sources and the scale of quantum gravity}
\author{Roberto Casadio}
\email{casadio@bo.infn.it}
\affiliation{Dipartimento di Fisica, Universit\`a di Bologna,
and I.N.F.N., Sezione di Bologna, via~Irnerio~46, 40126~Bologna, Italy}
\author{Remo Garattini}
\email{Remo.Garattini@unibg.it}
\affiliation{Facolt\`a di Ingegneria, Universit\`a di Bergamo,
viale~Marconi~5, 24044~Dalmine, and
I.N.F.N., Sezione di Milano, via~Celoria~16, Milano, Italy}
\author{Fabio Scardigli}
\email{fabio@phys.ntu.edu.tw}
\affiliation{Leung Center for Cosmology and Particle Astrophysics (LeCosPA),
Department of Physics, National Taiwan University, Taipei~106, Taiwan}
\affiliation{Yukawa Institute for Theoretical Physics, Kyoto University, Kyoto~606-8502, Japan}
\begin{abstract}
We review the General Relativistic model of a (quasi) point-like particle represented by
a massive shell of neutral matter which has vanishing total energy in the small-volume limit.
We then show that, by assuming a Generalised Uncertainty Principle, which implies
the existence of a minimum length of the order of the Planck scale, the total energy instead
remains finite and equal to the shell's proper mass both for very heavy and very light
particles.
This suggests that the quantum structure of space-time might be related to the classical
Equivalence Principle and possible implications for the late stage of evaporating black holes
are briefly mentioned. 
\end{abstract}
\maketitle
\section{Introduction}
In classical physics, nothing prevents us from describing
elementary particles as being point-like, except that the stress-energy tensor of
the electromagnetic and Newtonian gravitational fields diverge.
In fact, this problem can actually be avoided in General Relativity, as was first shown
in Ref.~\cite{adm60}, where localised sources were described as shells of matter
whose total energy remains finite in the ``point-like'' limit.
In particular, if the shell is electrically neutral, its total energy vanishes in
the small-volume limit (whose precise definition will be given in due course).
This result was interpreted as the fact that General Relativity does not
allow finite amounts of energy in a vanishingly small volume.
\par
There are however many reasons to believe that, in a quantum mechanical world,
the point-like limit is meaningless.
To begin with, one has the Heisenberg Uncertainty Principle to prevent complete
localisation in the phase space of Minkowskian theories.
Moving bottom up to a semiclassical scenario, in which gravity is still described
in terms of a background space-time, rigourous results and plausibility arguments
suggest the emergence of a Generalised Uncertainty Principle (GUP)~\cite{gup}.
The idea behind all proposed GUP's is that in a scattering experiment with
beams of energy $E$, the minimum accessible length is given by
\be
\delta x \geq \frac{\lp\,\mpl}{2\,E}+\frac{\alpha^2}{4}\,R_g(E)
\ ,
\label{dx}
\ee 
where $\alpha \simeq 1$~\footnote{Heisenberg's principle is recovered for
$\alpha\to 0$.
In Ref.~\cite{vagenas}, the equivalent parameter $\beta_0=\alpha^2$ was used.}
and $R_g$ is the gravitational radius associated
with the energy of the scattering process.
The latter is given by $R_g(E)=2\,\lp\,E/\mpl$
in the simplest approximation of the Schwarzschild geometry,
with $\lp$ and $\mpl$ the Planck length and mass
(we use units with $\hbar=c=1$ and the Newton constant $16\,\pi\,G_{\rm N}=\lp/\mpl$).
On minimising Eq.~\eqref{dx} with respect to $E$, one obtains
\be
E_{\rm min}=\alpha^{-1}\,{\mpl}
\ ,
\quad
\delta x_{\rm min}\equiv \lambda=\alpha\,\lp
\ .
\label{L}
\ee
Similar conclusions are also obtained from top-down approaches starting from
more fundamental theories, such as String Theory~\cite{string}
and Loop Quantum Gravity~\cite{loop}, which hint to space-time
non-commutativity~\cite{szabo} at short length scales.
From the phenomenological point of view, the existence of a minimum length
interestingly leads to universal corrections which might even be within the reach
of forthcoming experiments~\cite{vagenas}.
\par
We shall here review the neutral solution of Ref.~\cite{adm60}, also in
the more standard approach of Israel~\cite{israel}, and then investigate
what consequences follow from the existence of the minimum length $\lambda$
of Eq.~\eqref{L}.
\section{Classical shell model}
Following Ref.~\cite{adm60}, we consider the space-time generated by a shell
of bare mass $m_0$ and coordinate radius $r=\epsilon$.
\subsection{ADM~model}
For the interior ($0\le r<\epsilon$), we shall assume flat Minkowski space-time,
\be
\ud s^2_{\rm i}=-C^2\,\ud t^2+A^4\left(\ud r^2+r^2\,\ud\Omega^2\right)
\ ,
\ee
and for the exterior ($r>\epsilon$) the isotropic form of the Schwarzschild
metric~\cite{mtw},
\be
\ud s^2_{\rm o}=-\left(\frac{2r-M}{2r+M}\right)^2\!\!\ud t^2
+\left(1+\frac{M}{2r}\right)^4\!\!
\left(\ud r^2+r^2\,\ud\Omega^2\right)
,
\label{isoS}
\ee
where $M$, $C$ and $A$ are constants determined by the matching conditions at
$r=\epsilon$.
The metric~\eqref{isoS} is the well-known prototype of a wormhole, 
asymptotically flat both for $r\to\infty$ and $r\to 0$.
\par
The total energy of this spherically symmetric space-time with asymptotically
flat metric (for $r\to\infty$) is given by the surface integral~\cite{adm60,regge,visser}
\be
E=-\!\!
\lim_{R\to\infty}\left[
\int \frac{\ud\theta\,\ud\phi}{8\,\pi}\sqrt{g^{(2)}}\left(
K-K_0\right)_{r=R}
\right]
\ ,
\label{Eadm}
\ee
where $g^{(2)}$ and $K$ are, respectively, the determinant of the two-metric and the trace
of the extrinsic curvature of a two-sphere $^{2}S$ of radius $R$;
$K_0$ is the trace of the extrinsic curvature corresponding to embedding the
two-dimensional boundary $^{2}S$ in three-dimensional Euclidean space
and yields the Minkowski ``reference'' energy.
The curvature tensor $K_{ij}$ can be evaluated by introducing a Gaussian normal
coordinate $y$ such that $r(y=0)=R$ and the spatial part of the metric~\eqref{isoS}
reads
\be
\ud s^2_{(3)}=\ud y^2
+\left(1+\frac{M}{2\,r(y)}\right)^4r^2(y)\,\ud\Omega^2
\ .
\ee
One then finds
\be
K=2\,g^{\theta\theta}\,K_{\theta\theta}
=-\frac{\ud r}{\ud y}\,\frac{\partial g_{\theta\theta}}{\partial r}
\ ,
\ee
and
\be
E
&\!\!=\!\!&
-\!\!\lim_{R\to\infty}\left\{
\frac{1}{2}\left(1+\frac{M}{2\,R}\right)^{-2}\!\!
\frac{\partial}{\partial r}\!
\left[r^2\left(1+\frac{M}{2\,r}\right)^4
\right]_{r=R}
\!\!\!\!\!\!\!\!
+R
\right\}
\nonumber
\\
&\!\!=\!\!&
M
\ ,
\ee
which shows that $M$ is the Arnowitt-Deser-Misner (ADM) mass of the system,
as expected~\cite{adm60}.
\par
The shell matter at $r=\epsilon$ is represented by a $\delta^{(3)}$-function energy density,
\be
8\,\pi\,G_{\rm N}\,
\sqrt{g^{(3)}}\,T^t_{\ t}=-\frac{M_0}{2}\,\sqrt{\eta^{(3)}}\,\delta^{(3)}(r)
\ ,
\ee
where
\be
M_0=\frac{\lp\,m_0}{16\,\pi\,\mpl}
\ ,
\ee
$g^{(3)}$ is the determinant of the spatial metric, 
$\eta^{(3)}=r^4\,\sin^2\theta$, and
\be
4\pi\int_0^\infty \delta^{(3)}(r)\,r^2\,\ud r=1
\ ,
\label{d3}
\ee
with
\be
\delta^{(3)}(r)=0
\quad
{\rm for}\
|r-\epsilon|>\rho
\ ,
\ee
where $0<\rho\ll\epsilon$ (and the limit $\rho\to 0$ at the end of the computations
is understood).
\par
The relevant equation is given by the $tt$-component of the Einstein equations,
\be
\sqrt{\frac{g^{(3)}}{\eta^{(3)}}}\left(R^t_{\ t}-\frac{1}{2}\,R\right)
=\frac{\chi\,\nabla^2\chi}{4\,\pi}
=-\frac{M_0}{2}\,\delta^{(3)}(r)
\ ,
\label{G00}
\ee
where
\be
\chi=\left\{
\begin{array}{ll}
A
&
{\rm for}
\quad
0\le r<\epsilon
\\
\\
\strut\displaystyle\left(1+\frac{M}{2\,r}\right)
&
{\rm for}
\quad
r>\epsilon
\ ,
\end{array}
\right.
\label{chi}
\ee
and 
\be
\nabla^2\chi=r^{-2}\,\partial_r\left(r^2\,\partial_r\chi\right)
\ee
is the flat space Laplacian in spherical coordinates.
Continuity of the metric across the shell then implies
\be
A=1+\frac{M}{2\,\epsilon}
\ ,
\ee
and, upon integrating both sides of Eq.~\eqref{G00} in a spherical volume
around $r=\epsilon$, one obtains
\be
\frac{M_0}{2}
&\!\!=\!\!&
-\lim_{\rho\to 0}
\int\limits_{\epsilon-\rho}^{\epsilon+\rho}
\chi\,\frac{\partial}{\partial r}\left(r^2\,\frac{\partial\chi}{\partial r}\right)\,\ud r
\nonumber
\\
&\!\!=\!\!&
\lim_{\rho\to 0}\left\{
\int\limits_{\epsilon-\rho}^{\epsilon+\rho}
\left(\frac{\partial\chi}{\partial r}\right)^2 r^2\,\ud r
-\left[r^2\,\chi\,\frac{\partial\chi}{\partial r}\right]^{r=\epsilon+\rho}_{r=\epsilon-\rho}
\right\}
\nonumber
\\
&\!\!=\!\!&
-\left[\frac{r^2}{2}\,\frac{\partial}{\partial r}\left(1+\frac{M}{2\,r}\right)^2
\right]_{r=\epsilon}
\ .
\label{nchi}
\ee
One can finally write Eq.~\eqref{nchi} as~\cite{adm60}
\be
M=-\epsilon+\sqrt{\epsilon^2+2\,M_0\,\epsilon}
\ .
\label{M0}
\ee
which implies that $2\,M\sim\sqrt{M_0\,\epsilon}$ for $\epsilon\ll M_0$.
\begin{figure}[t!]
\centering
\raisebox{3.5cm}{$\frac{r}{M}$}
\epsfxsize=7cm
\epsfbox{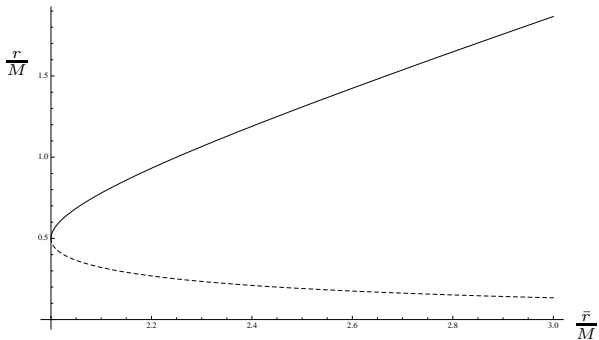}
$\frac{\bar r}{M}$
\caption{Radial coordinates $r_+$ (solid line) and $r_-$ (dashed line)
in Eq.~\eqref{rpm} as functions of areal radius $\bar r$ for fixed $M$.
\label{ADMSchw}}
\end{figure}
\subsection{Israel's junction equations}
The above result for the ADM mass can also be obtained from Israel's junction
equations for a static spherically symmetric shell~\cite{israel} in
Schwarzschild coordinates.
Let us define the usual areal radius ($\bar r\ge 0$) as
$\bar r=r$ for $r<\epsilon$ and
\be
\bar r=r\left(1+\frac{M}{2\,r}\right)^2
\ ,
\quad
{\rm for}\
r>\epsilon
\ .
\label{ar}
\ee
For $r>\epsilon$, we then find the two solutions
\be
r_\pm=\frac{1}{2}\left(\bar r-M\pm\sqrt{\bar r\left(\bar r-2\,M\right)}\right)
\ ,
\label{rpm}
\ee
which are both real for $\bar r>2\,M\ge 0$.
Note that $r=r_+(\bar r)$ and $r=r_-(\bar r)$ are
respectively increasing and decreasing in $\bar r$ (see Fig~\ref{ADMSchw}).
Therefore, $r_+>M/2$ spans the Schwarzschild manifold outside the horizon,
$\mathcal{M}^+=\{\bar r(r_+)>2M\}$,
whereas the region $0<r_-<M/2$ is a second copy of the same Schwarzschild manifold,
$\mathcal{M}^-=\{\bar r(r_-)>2M\}$.
The complete manifold $\mathcal{M}=\mathcal{M}^-\cup\mathcal{M}^+$
represents a ``wormhole'' whose ``throat'' has a minimum areal radius equal to
$2M$.
\par
We recall that wormhole metrics are usually given in the form~\cite{visser}
\be
\ud s_{w}^{2}=
e^{\strut\displaystyle{-2\phi(\bar r(x))}}\ud t^2+\ud x^2
+\bar{r}(x)^{2}\,\ud \Omega^{2}
\ ,
\ee
where
\be
\ud x=\frac{\sigma_\pm\,\ud\bar{r}}{\sqrt{1-b(\bar{r})/\bar{r}}}
\ ,
\ee
with $\sigma_+=+1$ in $\mathcal{M}^+$ and $\sigma_-=-1$ in $\mathcal{M}^-$;
$b\left(  \bar{r}\right)  $ is the ``shape function'' subjected to the
condition $b(\bar{r}_{t}) =\bar r_t$, where $\bar{r}_{t}=2\,M$ is
the throat corresponding to $x=0$, and $\phi(x)$ is the ``redshift
function''.
\par
The metrics inside and outside the shell can now be written as
\be
\ud s^2_{\rm i/o}=-f_{\rm i/o}\,\ud t^2+f^{-1}_{\rm i/o}\,\ud \bar r^2
+\bar r^2\,\ud\Omega^2
\ ,
\ee
with $f_{\rm i}=1$ and
\be
f_{\rm o}=1-\frac{b(\bar r)}{\bar r}=1-\frac{2\,M}{\bar r}
\ .
\ee
One of the junction equations for $\bar r>2\,M$ then reads~\cite{israel,ansoldi}
\be
M_0
&\!\!=\!\!&
\bar r(\epsilon)\left[\sqrt{f_{\rm i}}-\sigma_\pm\,\sqrt{f_{\rm o}}\right]_{\bar r=\bar r(\epsilon)}
\nonumber
\\
&\!\!=\!\!&
\bar r(\epsilon)-\sigma_\pm\,\sqrt{\bar r^2(\epsilon)-2\,M\,\bar r(\epsilon)}
\ ,
\label{M0i}
\ee
with $\sigma_+=+1$ for $\epsilon>M/2$ and $\sigma_-=-1$ for $0<\epsilon<M/2$.
The above expression exactly yields the relation~\eqref{M0} in isotropic coordinates
after using Eq.~\eqref{ar}.
In the Schwarzschild frame, it is also easier to see that Eq.~\eqref{M0i} requires that
$\bar r(\epsilon)$ remain finite for $\epsilon\to 0$ in order for $M_0$ to be finite
in $\mathcal{M}^-$ (where $\sigma_-=-1$).
This, together with the definition~\eqref{ar}, implies that $M$ must vanish
for $\epsilon\to 0$, again in agreement with Eq.~\eqref{M0}.
\par
Finally, the shell's surface tension is given by the second junction equation~\cite{israel,ansoldi},
\be
P=
\left.\frac{\partial M_0}{4\,\pi\,\partial\bar r^2}\right|_{\bar r(\epsilon)}
\!\!\!\!\!
=
\frac{1}{8\,\pi\,\bar r(\epsilon)}
\left[1
-\frac{\sigma_\pm\left[\bar r(\epsilon)-M\right]}
{\sqrt{\bar r^2(\epsilon)-2\,M\,\bar r(\epsilon)}}\right]
.
\label{P}
\ee
\begin{figure}[t!]
\centering
\epsfxsize=7cm
\epsfbox{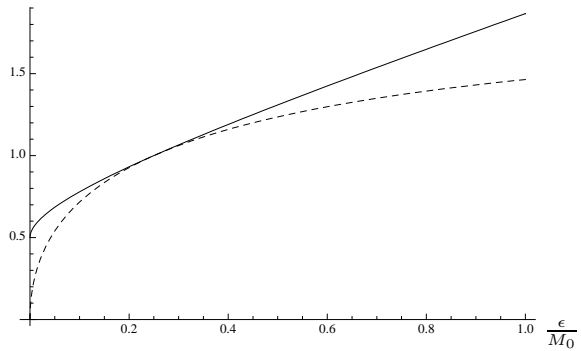}
$\frac{\epsilon}{M_0}$
\caption{Shell's areal radius ${\bar r}/{M_0}$ (solid line)
and throat radius ${2M}/{M_0}$ (dashed line) {\em vs\/}
shell's coordinate radius.
\label{r_eps}}
\end{figure}
\subsection{Small-volume limit}
Due to the dependence of $M$ on $\epsilon$, the shell areal radius
\be
\bar r(\epsilon)
=\frac{\epsilon}{4}\left(1+\sqrt{1+2\,\frac{M_0}{\epsilon}}\right)^2
\ ,
\label{reps}
\ee
is now a (single valued) monotonously increasing function of $\epsilon$
(see Fig.~\ref{r_eps}).
It also remains finite for $\epsilon\to 0$,
thus yielding a minimum shell volume $\propto \bar r^3(0)$ and area
$\propto \bar r^2(0)$ in the classical theory, where
\be
\bar r(0)=\frac{M_0}{2}
\label{csigma}
\ee
is of the order of the Schwarzschild radius of the mass $m_0$.
\par
This does not mean that the shell never enters the region $\mathcal{M}^-$
parameterized by $r_-$.
In fact, one finds that $\bar r(\epsilon)= 2M(\epsilon)$ for
\be
\epsilon=r_c\equiv \frac{M_0}{4}
\quad
{\rm and}
\quad
\bar r_c\equiv \bar r(r_c)=M_0=2\,\bar r(0)
\ ,
\ee
which only depend on $m_0$.
Thus, the shell is in $\mathcal{M}^+$ for $\epsilon>r_c$ and
in $\mathcal{M}^-$ for $0\le \epsilon<r_c$.
In the limit $\epsilon\to 0$, Eq.~\eqref{M0} yields $M(\epsilon)\to 0$, corresponding
to Minkowski space-time for $r>0$:
the throat pinches off and the shell becomes gravitationally inaccessible to observers
in $\mathcal M^{+}$.
From Eq.~\eqref{P} with $\sigma_-=-1$, one also sees that the surface tension remains
finite in this limit, namely
\be
P\simeq
\frac{1}{2\,\pi\,M_0}
\ .
\label{P0}
\ee
\section{Quantum theory}
\begin{figure}[t!]
\centering
\epsfxsize=7cm
\epsfbox{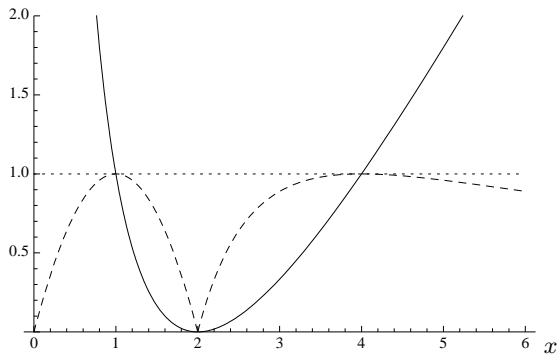}
$x$
\caption{Ratios $\gamma$ (solid line) and $\Gamma$ (dashed line) {\em vs\/}
rescaled shell's proper mass $m_0=16\,\pi\,\alpha\,\mpl\,x$.
Intersections of the two curves represent the shell exactly at $r=2\,M$.
\label{ratios}}
\end{figure}
If one views $m_0$ as the inertial mass and $M$ as the gravitational mass,
the result of Ref.~\cite{adm60} for neutral sources clearly conflicts with the
expectation that the Equivalence Principle (EP) holds for fundamental particles.
Let us then suppose that the shell is described by an effective quantum theory
with the minimum length $\lambda$ of Eq.~\eqref{L}.
\par
Given the existence of $\lambda$, the areal radius of a quantum-mechanically corrected
shell of bare mass $m_0$ should be bounded from below, that is
\be
\bar r(\epsilon)\gtrsim \bar r(\epsilon_\lambda)\equiv \bar r_\lambda\simeq \lambda
\ ,
\ee
and Eq.~\eqref{reps} provides a minimum value for its coordinate radius,
\be
\epsilon\gtrsim \epsilon_\lambda
\equiv
\lambda\left(1-\frac{M_0}{2\,\lambda}\right)^2
\equiv
\lambda\left(1-\frac{x}{2}\right)^2
\ .
\ee
Depending on the bare mass $m_0$ (or, equivalently, $x$), the shell will thus
be either in $\mathcal{M}^+$ or $\mathcal{M}^-$.
In particular, the shell is in $\mathcal{M}^-$ if the ratio
\be
\gamma\equiv\frac{\epsilon_\lambda}{r_c}
=\frac{4}{x}\left(1-\frac{x}{2}\right)^2
<1
\ ,
\ee
that is $1<x<4$, otherwise it is in $\mathcal{M}^+$.
In Fig.~\ref{ratios} we display both $\gamma$ and the ratio
\be
\Gamma\equiv \frac{2\,M_\lambda}{\bar r_\lambda}
=
\left\{
\begin{array}{ll}
\strut\displaystyle 2\,x\left(1-\frac{x}{2}\right)
&
\quad
{\rm if}\quad
0\le x<2
\\
\\
\strut\displaystyle \frac{8}{x}\left(1-\frac{2}{x}\right)
&
\quad
{\rm if}\quad
x>2
\ ,
\end{array}
\right.
\ee
where the vanishing total mass of the classical theory is also replaced by
the finite expression obtained for $\epsilon=\epsilon_\lambda$,
that is
\be
M_\lambda=
\left\{
\begin{array}{ll}
\strut\displaystyle M_0\left(1-\frac{M_0}{2\,\lambda}\right)
&
\quad
{\rm if}\quad
0\le M_0<2\,\lambda
\\
\\
\strut\displaystyle M_0\left(1-\frac{2\,\lambda}{M_0}\right)
&
\quad
{\rm if}\quad
M_0>2\,\lambda
\ .
\end{array}
\right.
\label{M0q}
\ee
\par
For an elementary particle and $\lambda=\alpha\,\lp$ (with $\alpha\gtrsim 1$),
it is natural to assume $x\sim m_0/\mpl\ll 1$.
One then finds that the shell is in $\mathcal{M}^+$ and its total energy is
\be
m_\lambda\equiv 16\,\pi\,\mpl\,\frac{M_\lambda}{\lp}
\simeq
{m_0}
\ .
\label{Ml}
\ee
Unlike in the purely classical theory, the EP therefore holds
for this quantum-mechanically corrected model of elementary particles, and 
does so regardless of the precise value of $\alpha\gtrsim 1$~\cite{vagenas}.
We also note that the shell's surface tension~\eqref{P} in this case
(with $\sigma_+=1$) is given by
\be
P_\lambda
\simeq
-\frac{\epsilon_\lambda}{\pi\,M_0^2}
\simeq
-\frac{\lambda}{\pi\,M_0^2}
\ .
\ee
\par
Further, for a mass $m_0\gg \mpl$ ($x\gg 1$), the shell is again
in $\mathcal{M}^+$ and its total energy (in units of length) is given by
Eq.~\eqref{Ml}.
The EP is thus preserved also for macroscopic objects, as
it should~\footnote{This behaviour of $M=M(M_0)$ for fixed $\lambda$
somehow resembles the R-duality in String Theory~\cite{string}.}.
\par
Significant corrections to the EP would only occur for $M_0\simeq 2\,\lambda$,
that is for particles with a mass around the Planck scale, which would 
therefore enter the region $\mathcal{M}^-$.
One such option would be micro-black holes, or black holes which have reached
the latest stages of their evaporation.
Eqs.~\eqref{M0q} and~\eqref{P} then predict vanishing total energy
$M\simeq 0$ for a remnant of proper mass $m_0\simeq 2\,\alpha\,\mpl$
and surface tension
\be
P\simeq \frac{1}{4\,\pi\,\alpha\,\lp}
\ .
\ee
\section{Conclusions}
We reviewed the classical description of elementary ``point-like'' neutral particles 
in General Relativity given long ago in Ref.~\cite{adm60} in terms of shells~\footnote{See
also Refs.~\cite{hsu} for more recent analyses.}.
A major findings in that paper was that neutral shells of finite ``bare'' mass $m_0$
have zero total energy in the small-volume limit and, therefore, they
do not interact gravitationally and violate the EP.
We applied a GUP type of argument, according to which this limit should be corrected so
as to admit a minimum length scale $\lambda$ of the order of the Planck length.
We then found that the energy of this ``point-like'' source is naturally equal to
$m_0$, regardless of the precise value of $\lambda$, and
deviations from the EP only occur for objects around the Planck size.
Such corrections support the possibility of evaporating black holes leaving
remnants of vanishing ADM~mass~\footnote{The remnants should have vanishing
specific heat~\cite{fabio}, like the non-commutative black holes of Ref.~\cite{nicolini}.}.
\par
It would be interesting to study what further implications our results
may have in the context of black hole quantum formation and
evaporation~\cite{ansoldi,qbh}, and the ultra-violet divergences of quantum
field theory~\cite{c}.
\end{document}